\documentstyle[11pt,aaspp4,psfig]{article}

\begin{document}     

\def\today{\ifcase\month\or
January\or February\or March\or April\or May\or June\or
July\or August\or September\or October\or November\or December\fi
\space\number\day, \number\year}


\baselineskip 0.210in

\newcommand{\squig}{$\sim$}
\newcommand{\squigleq}{\mbox{$^{<}\mskip-10.5mu_\sim$}}
\newcommand{\squiggeq}{\mbox{$^{>}\mskip-10.5mu_\sim$}}
\newcommand{\squiggeqmm}{\mbox{$^{>}\mskip-10.5mu_\sim$}}
\newcommand{\decsec}[2]{$#1\mbox{$''\mskip-7.6mu.\,$}#2$}
\newcommand{\decsecmm}[2]{#1\mbox{$''\mskip-7.6mu.\,$}#2}
\newcommand{\decdeg}[2]{$#1\mbox{$^\circ\mskip-6.6mu.\,$}#2$}
\newcommand{\decdegmm}[2]{#1\mbox{$^\circ\mskip-6.6mu.\,$}#2}
\newcommand{\decsectim}[2]{$#1\mbox{$^{\rm s}\mskip-6.3mu.\,$}#2$}
\newcommand{\decmin}[2]{$#1\mbox{$'\mskip-5.6mu.$}#2$}
\newcommand{\asecbyasec}[2]{#1$''\times$#2$''$}
\newcommand{\aminbyamin}[2]{#1$'\times$#2$'$}
\newcommand{\Lra}{$\Longrightarrow$}
\newcommand{\tmA}{\tablenotemark{a}}

\title{Serendipitous Discovery of a Cataclysmic Variable in the Globular
Cluster NGC\,6624\,\footnote{\ Based on observations with the NASA/ESA
Hubble Space Telescope, obtained at the Space Telescope Science Institute,
which is operated by the Association of Universities for Research in
Astronomy, Inc., under NASA contract NAS5-26555.}}
\author{Eric W. Deutsch, Bruce Margon, and Scott F. Anderson}
\affil{Department of Astronomy, 
       University of Washington, Box 351580,
       Seattle, WA 98195-1580\\
       deutsch@astro.washington.edu; margon@astro.washington.edu;
       anderson@astro.washington.edu}

\author{and\\ \vskip .1in Ronald A. Downes}
\affil{Space Telescope Science Institute,
       3700 San Martin Drive,
       Baltimore, MD 21218\\
       downes@stsci.edu}

\begin{center}
Accepted for publication in the Astronomical Journal\\
To appear in the 1999 December issue, Volume 118\\
{\it received 1999 July 28; accepted 1999 August 31}
\end{center}


\begin{abstract}

Despite indications that classical cataclysmic variable (CV) stars are
rare in globular clusters in general, and in the cluster NGC\,6624 in
particular, we have serendipitously discovered such a star $\sim6''$
from the cluster center. A {\it Hubble Space Telescope} spectrum of the
$m\sim22$ object shows strong, broad emission lines typical of numerous
field CVs, and the inferred optical and UV luminosity are also similar.
Our accidental observation also provides the first high-quality
ultraviolet spectrum of a globular cluster CV.  That we have detected
such an object in an observation that includes just a few percent of
the central area of the cluster may indicate that cluster CVs are
more common than previously thought, at least near the core.

\end{abstract}


\section{INTRODUCTION}

A small number of close binary stars are thought to dominate the dynamic
evolution of many globular star clusters (Hut et al. 1992, Bailyn 1995),
yet classes of such objects which are relatively easily found in the
field have proven frustratingly difficult to discover in clusters. A prime
example is cataclysmic variables (CVs), which call attention to themselves
via large amplitude light outbursts, and peculiar, ultraviolet-excess
colors in quiescence. With quiescent absolute magnitudes $M_V\sim7-8$,
modern ground-based photometric techniques should easily uncover
such objects in clusters with typical distances of $(m-M)\sim14$,
even with modest telescopes, unless all such objects are lost to the
crowded cores. Yet prior to the launch of {\it Hubble Space Telescope}
({\it HST}), we are aware of only two candidate identifications of CVs in
globular clusters, M5~V101 (Margon et al. 1981), whose classification and
cluster membership seems secure (Naylor et al. 1989, Shara et al. 1990),
and M30~V4 (Margon \& Downes 1983), whose membership is unclear (Shara
et al. 1990, Machin et al. 1991).

Observations from {\it HST} have certainly improved the situation,
although perhaps not as much as many would have expected. A few clusters
are now known to have a handful of spectroscopically-confirmed CVs;
a recent review is given by Grindlay (1999). However despite intensive
photometric and color-selected searches deep into the cores of a number
of clusters, few outbursting objects are found, and most authors believe
there is a serious discrepancy with theoretical predictions (Shara \&
Drissen 1995; Livio 1996; Shara et al. 1996, hereafter S96). Whether the
problem lies with formation/destruction rates, or some unique property
of cluster CVs, remains to be clarified, and will surely require a larger
sample of objects.

Here we discuss {\it HST} observations of the cluster NGC\,6624 using
the {\it Space Telescope Imaging Spectrograph} (STIS). This cluster
contains near its center a highly luminous bursting X-ray source with an
11-minute period, the shortest-known binary star; the system is thought to
be a double-degenerate (Stella et al. 1987, King et al. 1993, Anderson
et al. 1997). Another object in the cluster has also been suggested
by S96 as a candidate CV, although we discuss its nature further in \S
3.  As part of a program to study the central bright X-ray source,
we obtained deep STIS spectroscopic exposures of this object, and these
results will be discussed elsewhere (Deutsch et al. 2000). Here we report
the completely serendipitous discovery of $m\sim22$ emission line object,
which fell by good fortune in the STIS slit, and has the properties
of a classical CV. This object deepens the mystery of the CV content
of clusters.  S96 report a multi-epoch sensitive photometric search
of the core of NGC\,6624 which identifies only one candidate CV, and
suggest these objects are very rare, at least in this cluster. Yet we
have found another such an object completely by accident, in the first
long-slit ultraviolet spectral exposure made of the cluster.

\section{OBSERVATIONS AND ANALYSIS}

\subsection{STIS Spectra}

On 1998 March 14 we obtained 12150 s of integration in the center of
NGC\,6624 over 5 orbits with the {\it HST} STIS FUV-MAMA, using a \decsec{0}{5}
wide, $23''$ long slit.  The spectra are reprocessed and extracted using
CALSTIS 2.0 and the latest calibration files available as of 1999 April
20, where nearly half of the calibration files are updated from the
original pipeline processing.

In Fig. 1 we show a $10''$ long region of the cluster spectrum, which
has been background subtracted for display purposes.  The primary target,
the X-ray source hereafter denoted Star K, is clearly visible at the top.
Two other objects, labeled Stars 1 and 2, are visible.  No other objects
are discernible in the entire long-slit spectrum, including the region
not shown in Fig. 1.

The extracted spectrum of Star 1, binned $2\times$ to 1.17 \AA\
per bin, is displayed in Fig. 2.  Strong, broad emission lines
of N V $\lambda\lambda$1238,1242, Si IV $\lambda\lambda$1394,1403
and/or O IV] $\lambda$1401, C~IV~$\lambda\lambda$1548,1550, and He II
$\lambda$1640 are clearly detected.  The O I $\lambda$1304 feature is
most likely imperfectly subtracted geocoronal emission.  The location
of a few other common emission lines which may contribute to this
spectrum are also marked.  The continuum is very weakly detected
at $(0.4\pm0.2)\times10^{-17}$ erg cm$^{-2}$ s$^{-1}$ \AA$^{-1}$.
This spectrum is that of a classical cataclysmic variable in a quiescent
state (e.g., Wu et al. 1992).  We show in \S 3 that both the optical and
UV magnitudes of the object agree with those expected for a CV at the
distance of the cluster, and the probability of a chance superposition
of a non-member within $6''$ of the cluster core is small.  We conclude
that we have serendipitously discovered a CV in the cluster.

We have attempted to independently verify cluster membership via radial
velocity measurements of the spectrum, but the results are inconclusive.
We were forced to use an undesirably large slit width due to potential
problems in target acquisition in this complex field, and the resulting
configuration is poorly suited for absolute velocity determinations.
If we accept the nominal CALSTIS wavelength calibration, the inferred
radial velocity, obtained from several strong emission lines, is $-600$
km s$^{-1}$.  The uncertainty in this value is however dominated by
the zero point error introduced by the unknown location of the object
within the slit, which could be as large as $\pm 1,200$ km s$^{-1}$, so the
constraint is not meaningful. Further, we expect the radial velocity of
the star to vary, with unknown period and probably large amplitude, and so
cannot make a cogent estimate of the mean velocity after one observation.
We have also searched for radial velocity variations from orbit to orbit,
again with inconclusive results due to large uncertainties in the subsets
of the data.

The spectrum of Star 2 is a flat continuum of $(1.7\pm0.2)\times10^{-17}$
erg cm$^{-2}$ s$^{-1}$ \AA$^{-1}$, and featureless except
for possible detections of the common interstellar absorption
lines Si~II $\lambda1260$, C II $\lambda\lambda$1334,1335, and C
IV $\lambda\lambda$1548,1551.  When dereddened, the spectrum and
photometry are reasonably well described by a Kurucz model (Kurucz 1992)
of T$_{eff}=14$,000~K and $M_V\sim6.5$.  This is consistent with the
properties of the non-flickering ``NF'' objects discussed by Cool et
al. (1998) and Edmonds et al. (1999).

\subsection{WFPC2 and FOC Imagery}

We attempt to determine the precise location of the objects seen in
our spectroscopic observations by examining archival {\it HST} images.
We find F140W and F430W (pre-COSTAR) FOC images taken on 1992 August 13
which cover this field, as well as WFPC2 images taken on 1994 April 17
and 1994 October 15, obtained with a variety of filters.  There exist two
additional epochs of WFPC2 data, but as this field falls on the lower
resolution Wide Field Camera CCDs in those observations, they provide
no additional useful information.

Given the crowding in the cluster core, determination of which objects are
responsible for the observed spectra is not trivial.  Based on information
in the STIS header, the distance of Stars 1 and 2 from the bright Star
K is \decsec{4}{66} and \decsec{6}{34}, respectively, at a position
angle of $41^\circ$.  In Fig. 3 we show \asecbyasec{5}{5} regions of
the various FOC and WFPC2 images.  Overlaid are two arcs indicating the
calculated radii of these objects from Star K (not present in this field).
The nearly horizontal lines indicate the edges of a \decsec{0}{5} slit.
Therefore, we expect to find the objects responsible for the spectra near
the intersections of the arcs and the slit.  Residual uncertainty in the
slit angle is less than 1$^\circ$ and thus not important at this scale.
Finally, the figure assumes that Star K is centered on the slit.  If Star
K is in fact shifted slightly in the slit, then the location of the slit
we have drawn may be displaced up or down by up to \decsec{0}{25}; the
target acquisition scenario employed should place Star K very near the
center of the slit, however.

In the 1400 \AA\ FOC image, we find a faint object at the expected
location of Star 2, and label it Star B (recognizing that the association
is hardly guaranteed.)  We find no evidence for any detection at the
expected location of Star 1 in this F140W image.  Examining successively
longer wavelength images, we find that star B exhibits UV-excess between
the F336W and F439W images, but is hopelessly contaminated by light from
neighboring stars at longer wavelengths.  At the expected location of Star
1, we find a faint star which we label A at F439W and longer wavelengths,
although there is no evidence of it at F336W and F140W.  Star A is
positively detected at multiple epochs as well as multiple wavelengths,
although it falls just outside the FOC F430W image.
There is also another faint object \decsec{0}{15} NW of Star A, which is
also a plausible candidate.  No small adjustments in slit angle or slit
shift yield a better alignment.  In the astrometric frame of {\it HST}
image U2AS0101T, the coordinates for Star A are
  $\alpha(2000)={\rm18^h23^m}$\decsectim{40}{810},
  $\delta(2000)={\rm-30^\circ21'}$\decsec{35}{01}.
Although internally precise, this position has a probable uncertainty
of $\sim1-2''$ with respect to external frames.  Based on our estimate
of the cluster center in the same image, we find that Star A is
\decsec{6}{5}, or 1.8 core radii, from the center of NGC\,6624 (assuming
$r_c=\decsecmm{3}{6}$ from Harris 1996).  A detailed discussion of a
precise measurement of the cluster center, not relevant to our analysis,
is given by King et al. (1993).  Two radio pulsars are located within
$\sim10''$ of Star A (Biggs et al. 1994), but the positions are
definitely disjoint, given the quoted uncertainties.  In a cluster
of $r_c=\decsecmm{3}{6}$, it is perhaps not surprising that multiple
interesting but unrelated objects are in such proximity.

In Table 1 we present photometry for selected objects in the field.
We use a combination of profile-fitting and aperture photometry to derive
these magnitudes.  For WFPC2 images, aperture corrections are taken from
Table 2(a) in Holtzman et al. (1995b).  The photometric measurements
have not been corrected for geometric distortions, nor is any correction
for charge transfer efficiency losses (Holtzman et al. 1995b) applied;
for most of the images, these effects should contribute errors of only a
few percent.  We use the photometric zero points for the STMAG system from
Table 9 ($Z_{STMAG}$) in Holtzman et al. (1995a).  Approximate $1\sigma$
measurement uncertainties are also provided in the table.  Systematic
errors for all magnitudes due to uncertainties in detector performance and
absolute calibration are \squig 5\%.  Magnitude measurements presented
here are denoted $m_\lambda$, where $\lambda$ is the filter designation,
approximately indicating the central wavelength of the filter in nanometers.
In the STMAG system, zero points are set such that a flat spectrum
($f_\lambda={\rm constant}$) source will have identical magnitudes at
all wavelengths.

For the FOC measurements, we use the calibration provided in the header
of the images.  A STSDAS {\it synphot} calculation yields a calibration
value which differs by $\sim30$\%.  We use a Tiny TIM (Krist 1993)
synthetic profile to determine an aperture correction.  Since the F140W
images were obtained in an unusual mode and far UV calibration is often
difficult in any case, we suspect that the absolute calibration may not
be accurate to better than a factor of two.

In order to further check that the associations between Stars 1 and A,
and Stars 2 and B, are plausible, we estimate $m_{140}$ for Stars 1
and 2 at the time of the STIS observations by convolving the spectra
with the F140W bandpass and integrating the observed flux.  We estimate
$m_{140}=22.2$ for Star 1, and $m_{140}=20.9$ for Star 2, and estimate
uncertainties of 0.2 mag principally because the spectra do not fully
cover the F140W bandpass.  Given the various uncertainties in this
estimate and FOC calibration, the agreement between the STIS and FOC
(Table~1) magnitude estimates is excellent.  Aside from the observed level
of excitation in the spectrum, the lack of any brighter and/or spatially
extended image near the correct location for Star 1 provides confirmation
that the observed emission spectrum cannot be due to a cluster planetary
nebula or the background superposition of a low redshift AGN.  We conclude
that Star A is the likely source of the emission line spectrum.

We point out in passing yet another unusual object in the same field.
Star C (Table~1, Fig.~3) is extremely bright in these F140W images, but
a rather faint object in longer wavelength passbands.  When dereddened,
the $m_{140}$, $m_{336}$, and $m_{439}$ measurements are well fit by a
Kurucz model of T$_{eff}=35$,000~K.  This temperature is most sensitive to
the F140W measurement for which the calibration is poorest.  If the FOC
calibration is adjusted by 0.4 mag such that the FOC observed and STIS
convolved magnitudes are equal for Star B, the implied temperature for
Star C is T$_{eff}=30$,000~K.  This extraordinary UV excess is noteworthy
but otherwise not relevant to the present discussion, except as further
evidence of a multitude of exotic objects near the center of this cluster.

\bigskip
\section{DISCUSSION}

Our photometry in \S 2.2 together with the known distance and reddening of
the cluster permit a comparison of the luminosity of this object with that
of the far better-studied field CVs.  Adopting $(m-M)_0=14.50$ and
$E(B-V)=0.28$ (Harris 1996), our measured $m_{555}=21.8$ implies
$M_V\sim6.5$.  There are now four trigonometric parallaxes for classical
field CVs (Harrison et al. 1999, McArthur et al. 1999), which collectively
imply $<M_{V,min}\sim8.0>$, similar to the $<M_{V,min}\sim7.5>$ often quoted
from much larger samples (Warner 1995). Given the uncertainties in our
photometry for this faint object in a very crowded field, and our
single-epoch measurement of an undoubtedly variable star, the agreement
of the inferred luminosity of the new NGC\,6624 CV with those in the field is
gratifying. If the system is strongly magnetic, a possibility we consider
below, our observed magnitude is perhaps somewhat brighter than expected
from polars in the field, but, again given the uncertainties, not alarmingly
so.

We are aware of few if any ultraviolet spectra of globular cluster CVs.
Edmonds et al. (1999) display a low signal-to-noise spectrum of an object in
NGC~6397, where He\,II $\lambda$1640 is termed ``marginally detected" in
emission by those authors.  Through absolutely no credit to the current
authors, our spectrum is far better exposed.

The prominent He\,II $\lambda$1640 emission in our spectrum is deserving
of comment.  This line is normally not strong in classical CVs, but
is seen in polars.  Our spectrum is indeed quite similar to that of
AM~Her in this wavelength range (Greeley et al. 1999). Unfortunately
the simplest defining spectroscopic characteristic of AM~Her stars,
very strong He\,II $\lambda$4686 emission, is not accessible to
us. Although little can be inferred from one object, we note that
Grindlay et al. (1995) have suggested that magnetic white dwarfs
are preferentially produced in globular clusters (see also Grindlay
1999 and references therein).  These authors also discuss the issue of
possible confusion of the spectra of cluster CVs with those of quiescent
low-mass X-ray binaries (LMXBs), and those considerations also apply here.
Our spectrum does not unambiguously distinguish between the two cases;
for example, the quiescent LMXB Cen X-4 (V822~Cen) shows quite weak
He\,II $\lambda$4686 (McClintock \& Remillard 1980) and little or no
$\lambda$1640 emission (Blair et al. 1984), but Aql X-1 (V1333~Aql)
displays strong He\,II~$\lambda$4686 in quiescence (Garcia et al. 1999).
As more CVs than quiescent LMXBs are known in clusters, it seems most
conservative, but still uncertain, to continue the discussion of CVs.
As this object lies only a few arcsec from the brightest X-ray source
in any globular cluster, its X-ray properties are as yet unknown, but
may be accessible to observations from {\it Chandra}.

Although our detection of the ultraviolet continuum is weak, it is of
interest to ask if the observed ultraviolet flux agrees with that expected
from the field objects, as we have determined above is the case for the
visible band.  If we assume $A_{135\rm{nm}}=9\,E_{B-V}$ (Cardelli et al.
1989), our observed $f_{\lambda}(135)\sim0.5\times10^{-17}$
erg~cm$^{-2}$~s$^{-1}$~\AA$^{-1}$ corresponds to an extinction-corrected
monochromatic luminosity of $L_{\lambda}(135)\sim3\times10^{29}$
erg~s$^{-1}$~\AA$^{-1}$. For the quiescent U~Gem, for example,
$f_{\lambda}(135)\sim1.5\times10^{-13}$ erg~cm$^{-2}$~s$^{-1}$~\AA$^{-1}$
(Wu et al. 1992, Long et al. 1994, Long \& Gilliland 1999); this ratio
of 30,000 in observed fluxes speaks well to advances in ultraviolet
spectroscopic capabilities from the {\it International Ultraviolet
Explorer} ({\it IUE}) to the {\it HST} era. Adopting $d=100$~pc (Harrison et
al. 1999) and $E(B-V)=0.03$ (Panek and Holm 1984, Long \& Gilliland
1999) for U~Gem, we infer $L_{\lambda}(135)\sim2\times10^{29}$
erg~s$^{-1}$~\AA$^{-1}$ for that object. While this excellent level
of agreement is almost surely fortuitous, it reaffirms that the CV
in NGC\,6624, with the exception of its unusual environment, seems a
totally normal cataclysmic.  As various abnormal mass transfer scenarios
have been at times invoked to explain the lack of cluster CV outbursts,
this agreement is interesting.

There is one previous candidate for a CV in NGC\,6624, as discussed by
S96.  The identification was based on the presence of the object on two
consecutive {\it HST} images, but its absence in all other observations.
We have recovered the candidate in the archival {\it HST} images, and
derive its position as
  $\alpha(2000)={\rm18^h23^m}$\decsectim{41}{349},
  $\delta(2000)={\rm-30^\circ21'}$\decsec{56}{53}
using the astrometric header information in {\it HST} image U2KL0406T.
This position is $\sim25''$ from Star A, and thus this object cannot
be responsible for our spectrum.

Moreover, upon close examination of these data, we find that the
radial profile of this object is not compatible with a stellar one.
We suggest it is an image artifact of uncertain origin.  In Fig.~4,
we show a surface plot of the pixels around this object and a nearby,
typical star of similar total counts.  The S96 object has a profile
much too sharp as compared with this and other stars.  It is indeed
detected in two frames, and thus is not a charged particle hit, but
rather probably a group of hot pixels or calibration file defect.
We stress however that the removal of the S96 object from the list of
cluster CVs merely {\it strengthens} the primary scientific conclusion
of that paper, namely that erupting CVs are remarkably rare in clusters.

\section{CONCLUSION}

We have serendipitously discovered a mass-transfer close binary,
presumably with a degenerate companion, close to the core of NGC~6624. It
is likely that the system is a classical cataclysmic variable, although a
quiescent LMXB cannot be ruled out. The situation is a curious inversion
of the normal problem in globular cluster observations, where one
finds an interesting star but has difficulty obtaining the spectrum
due to severe crowding.  We have easily obtained the spectrum, but are
uncertain of which object is responsible for the emission, although we
advance as a reasonable candidate Star~A. The uncertainty in the exact
coordinates of the system does not however change the surprising result
that an object supposedly so rare has been found accidentally.

The object was found in a single STIS long-slit exposure encompassing
$\sim12$~arcsec$^2$ at an arbitrary position angle passing quite
close to the cluster center. A hypothetical observing program that
included all possible position angles with this slit, thus completely
mapping the cluster center to $r=3\,r_c$, would cover $\sim30\times$
more area. Tempered by the usual uncertainties of {\it a posteriori}
statistics, it seems quite likely that the center of NGC\,6624 contains
several, and possibly many, more CVs of the type we have discovered here.
If we now accept that these systems most certainly are present --- as
has indeed been theoretically expected for some time --- then one or
more mechanisms which suppress the expected outbursts and/or odd colors,
and thus hide globular cluster CVs, must certainly be operative.

\acknowledgments
\bigskip

Support for this work was provided by NASA through grant NAG5-7932, and
also grant AR-07990.01 from the ST\,ScI, which is operated by AURA, Inc.



\begin{deluxetable}{lrrrrrrrr}
\tablenum{1}
\tablecolumns{9}
\tablecaption{Photometry of Selected Objects in NGC\,6624}
\tablehead{
\colhead{Object} &
\colhead{$m_{140}$} &
\colhead{$\sigma$} &
\colhead{$m_{336}$} &
\colhead{$\sigma$} &
\colhead{$m_{439}$} &
\colhead{$\sigma$} &
\colhead{$m_{555}$} &
\colhead{$\sigma$}
}
\startdata
Star A \tmA  &$> 21.6$& \nodata & $> 23.0$& \nodata &  22.2 & 0.5  &    21.8 &     0.3 \nl
Star B       &   20.5 &    0.2  &    21.5 &    0.2  &  21.2 & 0.3  & $> 21.5$& \nodata \nl
Star C       &   19.2 &    0.1  &    21.3 &    0.2  &  21.8 & 0.3  & $> 21.5$& \nodata \nl
\enddata
\tablenotetext{a}{\,suggested identification of the
spectroscopically-discovered cataclysmic variable}
\end{deluxetable}

\clearpage

\begin{figure}
\plotone{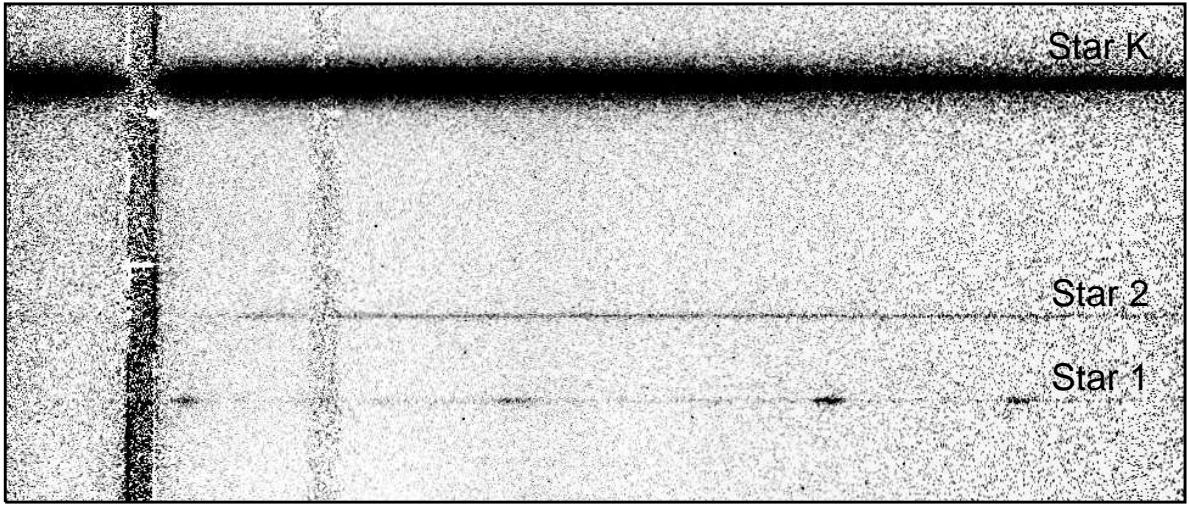}
\caption{550 \AA\ $\times\ 10''$ {\it HST} STIS long slit spectrum of the
core of NGC\,6624.  The low-mass X-ray binary optical counterpart, Star K,
is most prominent at the top.  Star 1 is a cataclysmic variable with broad
emission lines.  Star 2 is a UV-excess, featureless object also discussed
in the text.  The background has been subtracted in this image and the
two vertical bands are residuals from subtracted geocoronal Ly $\alpha$
and O I $\lambda 1304$ emission.  The N V $\lambda\lambda$1238,1242,
Si IV $\lambda\lambda$1394,1403, C IV $\lambda\lambda$1548,1550, and He
II $\lambda$1640 emission in Star~1 is clearly visible.}
\end{figure}

\begin{figure}
\plotone{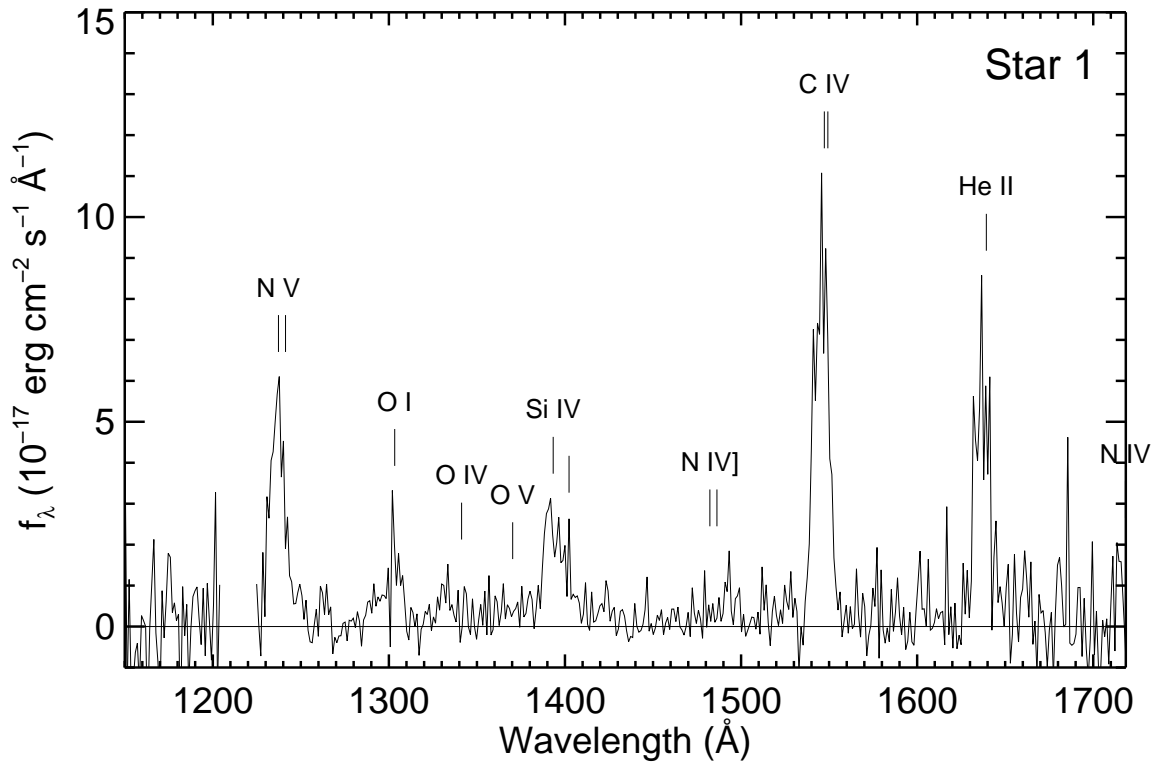}
\caption{The extracted, flux-calibrated {\it HST} STIS spectrum for
the newly discovered cataclysmic variable in NGC\,6624.  Geocoronal Ly
$\alpha$ badly mars the spectrum and has been deleted; the residual O
I $\lambda 1304$ emission is probably also geocoronal.}
\end{figure}

\begin{figure}
\hskip0.7in\psfig{file=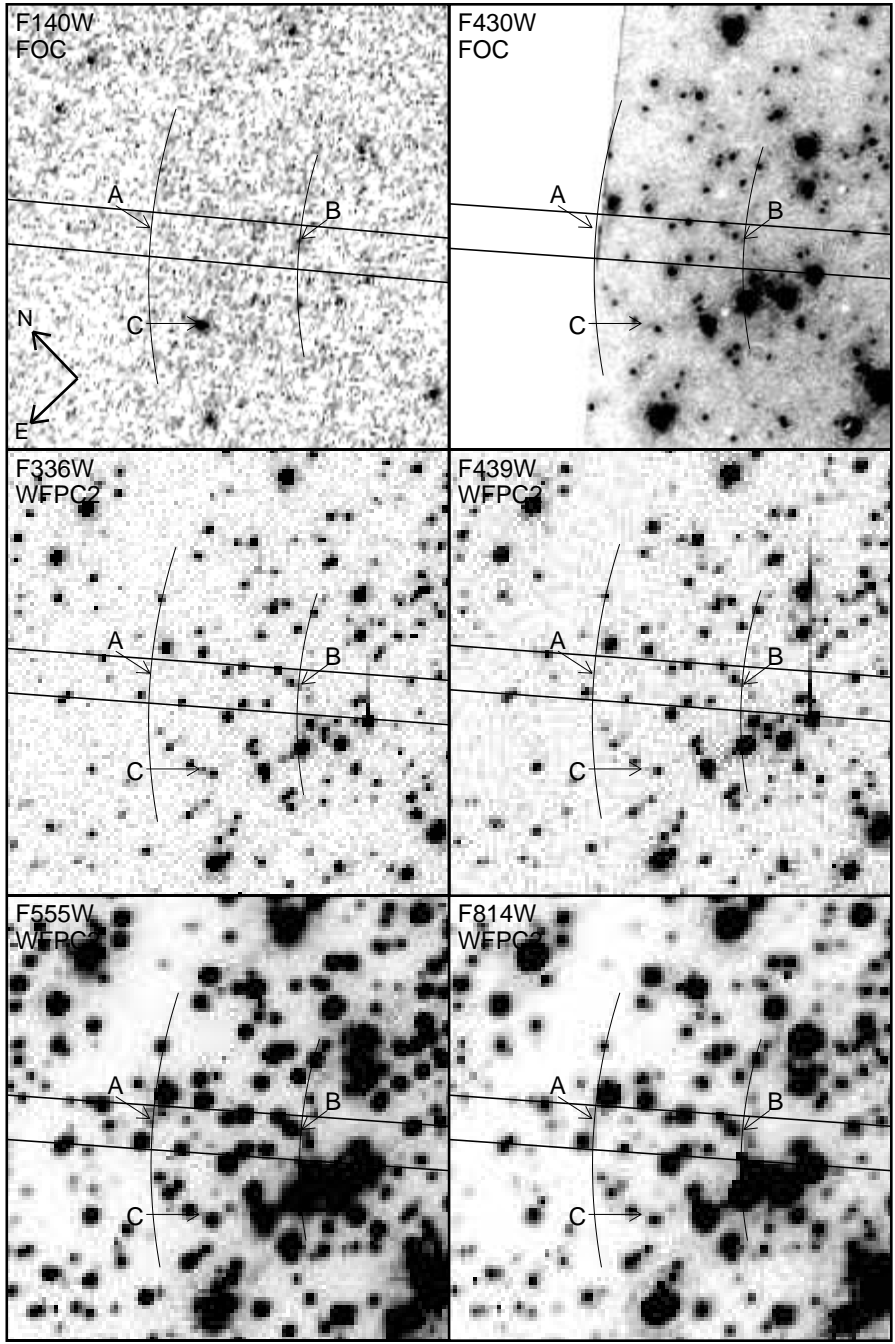,height=7in}
\caption{\asecbyasec{5}{5} field near the location of the STIS spectra
in NGC\,6624.  Each left/right pair of archival exposures was obtained at
the same epoch.  The two arcs represent the distances of our two spectra
from Star K (not itself in this field).  The two nearly horizontal lines
represent the STIS \decsec{0}{5} slit; thus the spectra discussed here
must arise from objects on or very near the arcs, and within the slit.
Note the rotated cardinal direction.  Various objects discussed in the
text are labeled.  We suggest that Star A is the optical counterpart to
the cataclysmic variable spectrum displayed in Fig.~2.}
\end{figure}

\begin{figure}
\hskip0.7in\psfig{file=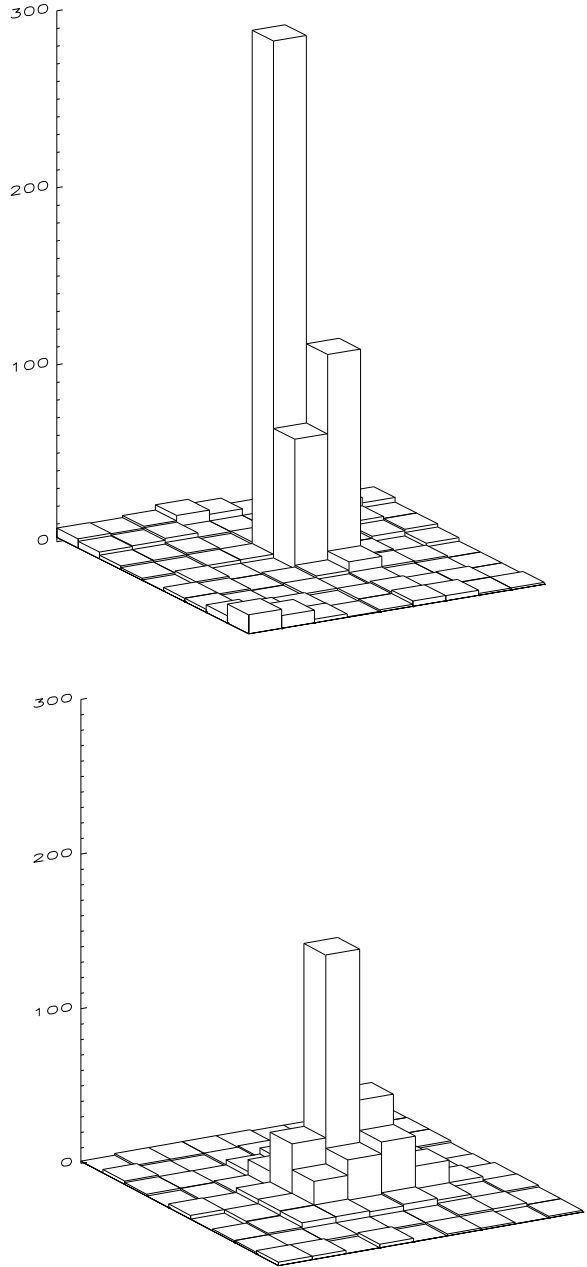,height=7in}
\caption{({\it top}) Surface plot of the Shara et al.\ (1996) NGC\,6624
dwarf nova candidate.  Each block is one individual pixel on the {\it HST}
Wide Field Camera detector image U2KL0406T.  ({\it bottom}) Surface plot
of a nearby star which has similar total flux within a 3 pixel radius.
The Shara et al.\ object is much sharper than a normal stellar profile,
and likely an image artifact.}
\end{figure}

\end{document}